\documentstyle[11pt]{article}

\font\msbm=msbm10 at 11pt
\font\eufm=eufm10 at 11pt

\font\eufma=eufm8 at 8pt
\font\msbma=msbm8 at 8pt

\newtheorem{proposition}{Proposition}

\newcounter{propapp}
\newcommand{\propappc}{\setcounter{propapp}{\value{proposition}}%
\renewcommand{\theproposition}{\Alph{proposition}}}
\newcounter{eqnapp}
\newcommand{\eqnappc}{\setcounter{eqnapp}{\value{equation}}%
\renewcommand{\theequation}{\mbox{A\arabic{eqnapp}}}%
\stepcounter{eqnapp}}

\newcommand{\DS}{\displaystyle}
\newcommand{\BD}{\begin{displaymath}}
\newcommand{\ED}{\end{displaymath}}
\newcommand{\BE}{\begin{enumerate}}
\newcommand{\EE}{\end{enumerate}}
\newcommand{\BA}{\begin{eqnarray*}}
\newcommand{\EA}{\end{eqnarray*}}
\newcommand{\BG}{\begin{equation}}
\newcommand{\EG}{\end{equation}}
\newcommand{\BGG}{\begin{eqnarray}}
\newcommand{\EGG}{\end{eqnarray}}

\newcommand{\lp}{\left (}
\newcommand{\rp}{\right )}

\newcommand{\lbr}{\left \{}
\newcommand{\rbr}{\right \}}

\newcommand{\Comug}[2]{\left [#1,#2\right ]_{g}}

\newcommand{\AComug}[2]{\left [#1,#2\right ]_{g}^{+}}
\newcommand{\Poisson}[3]{\left\{#1,#2\right\}^{#3}}
\newcommand{\Poissong}[3]{\left\{#1,#2\right\}_{g}^{#3}}

\newcommand{\Over}[1]{\overline{#1}}
\newcommand{\Under}[1]{\underline{#1}}

\newcommand{\Rom}[1]{{\rm #1}}
\newcommand{\Cal}[1]{{\cal #1}}

\newcommand{\Oger}[1]{\mbox{{\eufm #1}}}
\newcommand{\Ogera}[1]{\mbox{{\eufma #1}}}

\newcommand{\Ams}[1]{\mbox{{\msbm #1}}}
\newcommand{\Amsa}[1]{\mbox{{\msbma #1}}}

\newcommand{\CC}{\mbox{{\msbm C}}}

\newcommand{\NN}{\mbox{{\msbm N}}}
\newcommand{\ZZg}{\mbox{{\msbm Z}$_{2}$}}
\newcommand{\NNN}{\mbox{{\msbm N}$_{0}$}}
\newcommand{\CCa}{\mbox{{\msbma C}}}

\newcommand{\NNNa}{\mbox{{\msbma N}}_{0}}

\newcommand{\GMatk}[3]{\Ams{M}(#1\vert #2;{\msbm #3})}
\newcommand{\GMatkg}[4]{\Ams{M}(#1\vert #2;{\msbm #3})_{\Over{#4}}}

\newcommand{\Mat}[1]{\Ams{M}(#1)}
\newcommand{\Matnm}[2]{\Ams{M}(#1,#2)}
\newcommand{\GMat}[2]{\Ams{M}(#1\vert #2)}
\newcommand{\GMatg}[3]{\Ams{M}(#1\vert #2)_{\Over{#3}}}
\newcommand{\GMatnm}[4]{\Ams{M}(#1\vert #2,#3\vert #4)}
\newcommand{\GMatnmg}[5]{\Ams{M}(#1\vert #2,#3\vert #4)_{\Over{#5}}}

\newcommand{\GMata}[2]{\Amsa{M}(#1\vert #2)}

\newcommand{\Perm}[1]{\mbox{{\eufm S}}_{#1}}
\newcommand{\Perma}[1]{\mbox{{\eufma S}}_{#1}}

\newcommand{\GGl}[2]{\mbox{{\eufm gl}}(#1\vert #2)}
\newcommand{\GGlg}[3]{\mbox{{\eufm gl}}(#1\vert #2)_{\Over{#3}}}
\newcommand{\GSl}[2]{\mbox{{\eufm sl}}(#1\vert #2)}
\newcommand{\GSla}[2]{\mbox{{\eufma sl}}(#1\vert #2)}
\newcommand{\GSlg}[3]{\mbox{{\eufm sl}}(#1\vert #2)_{\Over{#3}}}

\newcommand{\Sl}[1]{\mbox{{\eufm sl}}(#1)}
\newcommand{\Gl}[1]{\mbox{{\eufm gl}}(#1)}

\newcommand{\GInd}[3]{\Oger{I}^{#1\vert #2}_{#3}}
\newcommand{\GInda}[3]{\Ogera{I}^{#1\vert #2}_{#3}}

\newcommand{\End}[1]{{\rm End}(#1)}

\newcommand{\Cent}[1]{{\cal Z}(#1)}
\newcommand{\GCent}[1]{{\cal Z}^{g}(#1)}
\newcommand{\GCentg}[2]{{\cal Z}^{g}(#1)_{\Over{#2}}}

\newcommand{\Homp}[3]{{\rm Hom}^{#1}(#2;#3)}

\newcommand{\Der}[1]{\Oger{Der}(#1)}
\newcommand{\GDer}[1]{\Oger{Der}^{g}(#1)}
\newcommand{\GDerg}[2]{\Oger{Der}^{g}(#1)_{\Over{#2}}}

\newcommand{\GHam}[1]{\Oger{Ham}^{g}(#1)}

\newcommand{\Cu}{${\cal C}^{\infty}$}
\newcommand{\Cuo}[1]{{\cal C}^{\infty}(#1)}

\newcommand{\GVecf}[1]{\Oger{V}^{g}(#1)}

\newcommand{\Diff}[1]{\Omega(#1)}
\newcommand{\Diffp}[2]{\Omega^{#1}(#2)}
\newcommand{\GDiff}[1]{\Omega^{g}(#1)}
\newcommand{\GDiffp}[2]{\Omega^{g,#1}(#2)}

\newcommand{\GDiffpg}[3]{\Omega^{g,#1}(#2)_{\Over{#3}}}

\newcommand{\DiffC}[1]{\Omega_{\Cal{Z}}(#1)}

\newcommand{\GDiffC}[1]{\Omega^{g}_{\Cal{Z}^{g}}(#1)}
\newcommand{\GDiffpC}[2]{\Omega^{g,#1}_{\Cal{Z}^{g}}(#2)}

\begin{document}

\pagenumbering{arabic}
\vspace*{0.5cm}

$\qquad \qquad \qquad \qquad \qquad \qquad \qquad \qquad \qquad \qquad \qquad \qquad \qquad \qquad \qquad \qquad$ UWThPh-6-1999

\vspace{0.7cm}

\LARGE
\centerline{{\bf Graded Differential Geometry of}}

\centerline{{\bf Graded Matrix Algebras}}

\vspace{1cm}

\Large
\centerline{H.Grosse$^{\rm{a},}$\footnote{Tel.: +43 1 31367 3413, Fax:
+43 1 317 2220, E-mail: grosse@doppler.thp.univie.ac.at}
and G.Reiter$^{\rm{a},\rm{b},}$\footnote{Research partly supported by the
``Fonds zur F\"orderung der wissenschaftlichen Forschung (FWF)''
through the research project P11783-PHY ``Quantum Field Theory on Noncommutative Manifolds''.}$^{,}$\footnote{Tel.: +43 316 873 8670, Fax:
+43 316 873 8678, E-mail: reiter@itp.tu-graz.ac.at}}

\vspace{7mm}

\normalsize
\centerline{\parbox{115mm}
{$^{\rm{a}}$ Universit\"{a}t Wien, Institut f\"ur Theoretische Physik, Boltzmanngasse 5,
A-1090 Wien, Austria}}

\vspace{2mm}

\centerline{\parbox{115mm}
{$^{\rm{b}}$ Technische Universit\"{a}t Graz, Institut f\"ur Theoretische Physik, Petersgasse 16, A-8010 Graz, Austria}}

\vspace{1cm}

\large
\centerline{{\bf Abstract}}
\normalsize

\vspace{4mm}

\centerline{\parbox{140mm}
{We study the graded derivation-based noncommutative differential geometry of
the $\ZZg$-graded algebra $\GMat{n}{m}$ of complex $(n+m)\times(n+m)$-matrices with the
``usual block matrix grading'' (for $n\neq m$). Beside the (infinite-dimensional) algebra of graded forms the graded Cartan calculus, graded symplectic structure, graded vector bundles,
graded connections and curvature are introduced and investigated. 
In particular we prove the universality of the graded derivation-based first-order
differential calculus and show, that $\GMat{n}{m}$ is a ``noncommutative graded manifold'' in a stricter sense: There is a natural body map and the cohomologies of $\GMat{n}{m}$ and its body coincide (as in the case of ordinary graded manifolds). 
\begin{tabbing}
\> geometry, matrix geometry
\end{tabbing}}}

\section{Introduction}

The basic idea of noncommutative geometry \cite{Connes1}, that is the formulation of differential geometric concepts on more general algebras than the algebras of \Cu-functions on differentiable manifolds, is at least conceptionally rooted in the fact, that all the information about the differentiable manifold and its sheaf of differentiable functions is encoded in its algebra of global \Cu-functions such that differential geometry can be formulated in terms of the latter algebras. 
Although the $\ZZg$-graded algebra of global sections of the structure sheaf of a graded manifold is a ``baby-noncommutative geometry'' the differential geometry of graded manifolds is treated and interpreted in the spirit of classical differential and algebraic geometry.
So graded manifolds should not be seen as specific noncommutative geometries to which
the general methods of noncommutative geometry applies but rather as a conceptional starting
point of a ``super-generalization'' of noncommutative geometry. Because graded manifolds
are completely determined by the $\ZZg$-graded algebra of global sections of their structure sheafs \cite{Bartocci1} the natural class of objects to which such a generalization applies
are $\ZZg$-graded real respectively complex algebras.\newline
There exist already several articles and books in the literature dealing with various 
aspects of \ZZg-graded \CC-algebras, supersymmetry and noncommutative geometry. 
Without being complete let us just mention \cite{Kastler1,Pittner1}, where notions as cyclic cohomology and Fredholm modules are treated in the \ZZg-graded setting, \cite{Frohlich1}, where supersymmetry is employed to establish metric, K\"a{h}ler and symplectic structures 
in noncommutative geometry, \cite{Kalau1}, where the concept of a spectral triple is extended to algebras which contain bosonic and fermionic degrees of freedom and 
\cite{Dubois-Violette2,Kerner1}, where a possibility of generalizing matrix geometry to the $\ZZg$-graded framework is presented. Here we want to adopt a somewhat different point of view. \newline
If $\Cal{O}(\Rom{X})$ is the $\ZZg$-graded algebra of global sections of the (complexified) structure sheaf of some graded manifold, (complex) global graded vector fields on
the graded manifold are by definition graded derivations of $\Cal{O}(\Rom{X})$. All
global graded vector fields $\GVecf{\Rom{X}}$ form a complex Lie subsuperalgebra and a
graded module over the graded center $\GCent{\Cal{O}(\Rom{X})}$.
(Complex) global graded $p$-forms for $p\in\NN$ are defined as $p$-fold 
$\GCent{\Cal{O}(\Rom{X})}$-graded-multilinear, graded-alternating maps from $\GVecf{\Rom{X}}$
to $\Cal{O}(\Rom{X})$ and one can form the $\NNN\times\ZZg$-bigraded \CC-vector space 
$\GDiff{\Rom{X}}$ of global graded forms as direct sum of all $\ZZg$-graded \CC-vector spaces
$\GDiffp{p}{\Rom{X}}$ of global graded $p$-forms. The graded wedge product as well as the
whole graded Cartan calculus on $\GDiff{\Rom{X}}$ can be introduced (see \cite
{Bartocci1,Kostant1} for example) by employing only the facts, that $\GVecf{\Rom{X}}$ is a $\CC$-Lie superalgebra and $\ZZg$-graded $\GCent{\Cal{O}(\Rom{X})}$-module and that $\Cal{O}(\Rom{X})$ is a $\ZZg$-graded $\CC$-algebra.\newline
The important feature of the recipe for the construction of the graded deRham complex and the graded Cartan calculus formulated above is, that it uses only the graded algebra structure
of $\Cal{O}(\Rom{X})$. In particular it does not use the graded commutativity of
$\Cal{O}(\Rom{X})$ and we can define on arbitrary $\ZZg$-graded \CC-algebras noncommutative
graded differential calculi.\newline
What we have just described is mutatis mutandis the basic idea of the so-called 
derivation-based differential calculi \cite{Djemai1,Dubois-Violette1,Dubois-Violette5} transposed to the $\ZZg$-graded setting. Such graded derivation-based differential calculi were investigated for arbitrary, but graded-commutative $\ZZg$-graded algebras in the framework of $\ZZg$-graded Lie-Cartan pairs \cite{Jadczyk1,Kastler1,Pittner1}.

Motivated by the rich differential geometric structure of ordinary matrix algebras 
\cite{Djemai1,Dubois-Violette1,Madore10} and by the our previous investigation of the fuzzy supersphere \cite{Grosse1,Grosse15}, where each truncated supersphere was a graded matrix algebra in particular, we will investigate especially the differential calculus based on all graded derivations on the $\ZZg$-graded \CC-algebra $\GMat{n}{m}$ of complex $(n+m)\times(n+m)$-matrices $(n,m\in\NNN,n+m\in\NN)$, $\ZZg$-graded by declaring
the vector subspace
\BG
\label{i1}
\GMatg{n}{m}{0} := \lbr
M=\lp\begin{array}{ll}
M_{1} & 0 \\
0 & M_{4} \\
\end{array}
\rp
\Big\vert~M_{1}\in\Mat{n}, M_{4}\in\Mat{m}~\rbr 
\EG
of $\GMat{n}{m}$ as even and the vector subspace
\BG
\label{i2}
\GMatg{n}{m}{1} := \lbr
M=\lp\begin{array}{ll}
0 & M_{2} \\
M_{3} & 0 \\
\end{array}
\rp
\Big\vert~M_{2}\in\Mat{n,m}, M_{3}\in\Mat{m,n}~\rbr 
\EG
of $\GMat{n}{m}$ as odd. Here $\Mat{n,m}$ and $\Mat{n}$ denote the vector space of
$n\times m$- respectively the algebra of $n\times n$-matrices and we will always assume $n\neq m$.\newline
Chapter 2 is devoted to the precise definition of the graded derivation-based differential
calculus on $\GMat{n}{m}$ as described above and its immediate consequences. The resulting
complexes are nothing else than the complexes of Lie superalgebra cohomology with
values in $\GMat{n}{m}$ and typically infinite. The latter fact shows in particular,
that the complex is completely different to that proposed in \cite{
Dubois-Violette2,Kerner1}.\newline
In chapter 3 we continue the investigation of the differential calculus using the facts, that
there exist graded-commutative homogeneous bases in the $\ZZg$-graded $\GMat{n}{m}$-modules
of all graded $p$-forms and that all graded derivations of $\GMat{n}{m}$ are inner. Especially
we construct an invariant graded 1-form, which determines the differential of graded
matrices in terms of graded commutators (within the graded algebra of graded forms) and
show, that the first-order differential calculus is universal.\newline
Associated with every graded manifold there exists an even, surjective algebra
homomorphism $\beta_{\Rom{X}}$ from the $\ZZg$-graded algebra of global sections of the (complexified) structure sheaf of the graded manifold $\Cal{O}(\Rom{X})$ 
to the algebra $\Cuo{\Rom{X}}$ of (complex) \Cu-functions on its body manifold $\Rom{X}$. We
call this map, which is the key to all further developments in graded manifold theory, the body map. In chapter 4 we show, that there exists a natural noncommutative analogue to the body map. It induces an isomorphism between the graded derivation-based cohomology
of $\GMat{n}{m}$ and the derivation-based cohomology of its body, such that the situation described by a theorem of Kostant \cite{Kostant1} is generalized to the noncommutative 
case.\newline
In chapter 5 we study the graded symplectic geometry of $\GMat{n}{m}$. As for ordinary matrix geometry \cite{Djemai1,Dubois-Violette1}, which is included as special case, there exists a graded symplectic structure such that the induced graded Poisson bracket on $\GMat{n}{m}$ is ($i$ times) the graded commutator on $\GMat{n}{m}$.\newline
In the last chapter we investigate the noncommutative generalization of graded vector
bundles over graded manifolds. Graded vector bundles over a graded manifold 
$(\Rom{X},\Cal{O})$ are usually introduced as locally graded-free $\Cal{O}$-modules
\cite{Bartocci1,Hernandez1,Kostant1}. In the spirit of noncommutative geometry \cite{Connes1,Gracia-Bondia1} we concentrate on the module of 
global sections and introduce graded vector bundles over $\GMat{n}{m}$ as $\ZZg$-graded, finitely generated (graded-projective) $\GMat{n}{m}$-modules. Concepts like connections and curvature can be generalized to the $\ZZg$-graded noncommutative setting.\newline
In addition we have included an appendix in which we analyze the associative product
of supertrace-free, graded matrices. The results of this analysis are used for a 
minimality proof in chapter 3. 

There will appear lots of \ZZg-graded objects. If the object is denoted by ${\cal A}$ its
even part is denoted by ${\cal A}_{\Over{0}}$, its odd part by ${\cal A}_{\Over{1}}$.
If $a$ is some homogeneous element of such an object its degree will be denoted by $\Over{a}$. 
Speaking of grading in the context of an ungraded object we mean, that the object is endowed
with its trivial graduation. If for some construction the $\ZZg$-grading is indicated by
an index ``$g$'' we omit this index in the case of trivial graduation.

\section{Graded derivation-based differential calculus on graded matrix algebras}

We will interpret the \CC-Lie superalgebra and $\ZZg$-graded $\GCent{\GMat{n}{m}}$-module
$\GDer{\GMat{n}{m}}$ of all graded derivations of $\GMat{n}{m}$ as ```noncommutative
graded vector fields'' on $\GMat{n}{m}$. Because $\GMat{n}{m}$ is graded-central, 
\BG 
\label{d1}
\GCent{\GMat{n}{m}} = \GCentg{\GMat{n}{m}}{0} = \CC\,1_{n+m} \cong \CC \, , 
\EG
the concept of graded $\GCent{\GMat{n}{m}}$-multilinearity reduces to ordinary 
$\CC$-multilinearity and we can employ ideas and results of Lie superalgebra cohomology
(see \cite{Fuks1,Scheunert8}) for the construction of the graded derivation-based 
differential calculus on $\GMat{n}{m}$.

For every natural number $p\in\NN$ let us denote by $\Homp{p}{\GDer{\GMat{n}{m}}}
{\GMat{n}{m}}$ the $\ZZg$-graded \CC-vector space of all $p$-linear maps $\GDer{\GMat{n}{m}}\times\stackrel{p}{\ldots}\times\GDer{\GMat{n}{m}}
\longrightarrow\GMat{n}{m}$ and by $\Perm{p}$ the symmetric group of $p$ letters. 
Introducing a commutation factor 
$\gamma_{p}:\Perm{p}\times\ZZg\times\stackrel{_p}{\ldots}\times\ZZg\longrightarrow\{\pm 1\}$
via
\BG
\label{d2}
\gamma_{p}(\sigma;\Over{i}_{1},\cdots,\Over{i}_{p}) :=
\prod_{r,s=1,\cdots,p;r<s\atop
\sigma^{-1}(r)>\sigma^{-1}(s)}
(-1)^{\Over{i}_{r}\Over{i}_{s}} \, ,
\EG
we can define a representation $\pi$ of $\Perm{p}$ on $\Homp{p}{\GDer{\GMat{n}{m}}}
{\GMat{n}{m}}$ by
\BG
\label{d3}
\lp\pi_{\sigma}\omega\rp(D_{1},\cdots,D_{p}) :=
\gamma_{p}(\sigma;\Over{D}_{1},\cdots,\Over{D}_{p})
\omega(D_{\sigma(1)},\cdots,D_{\sigma(p)})
\EG
for all $\omega\in\Homp{p}{\GDer{\GMat{n}{m}}}{\GMat{n}{m}}$, all homogeneous $D_{1},\cdots,D_{p}\in\GDer{\GMat{n}{m}}$ and all $\sigma\in\Perm{p}$. Now by definition a $p$-linear map $\omega\in\Homp{p}{\GDer{\GMat{n}{m}}}{\GMat{n}{m}}$ is called 
graded-alternating if
\BG
\label{d4}
\pi_{\sigma}\omega = \Rom{sgn}\sigma\,\omega
\EG
is fulfilled for all $\sigma\in\Perm{p}$ and we interpret such maps as graded $p$-forms on 
$\GMat{n}{m}$. All graded $p$-forms on $\GMat{n}{m}$ form a graded vector
subspace of $\Homp{p}{\GDer{\GMat{n}{m}}}{\GMat{n}{m}}$, which we will denote by
$\GDiffp{p}{\GMat{n}{m}}$.\newline 
A general graded form on $\GMat{n}{m}$ is an element of the direct sum 
\BG
\label{d5}
\GDiff{\GMat{n}{m}} := \bigoplus_{p\in\NNNa}\GDiffp{p}{\GMat{n}{m}} \, ,
\EG
where we set $\GDiffp{0}{\GMat{n}{m}}:=\GMat{n}{m}$.
Employing the multiplicative structure of $\GMat{n}{m}$ we can proceed exactly as in the case
of graded manifolds \cite{Bartocci1,Kostant1} (respectively graded Lie-Cartan pairs \cite{Jadczyk1,Pittner1}) to introduce a graded wedge product on $\GDiff{\GMat{n}{m}}$. So we first define for all $p,p'\in\NNN,\Over{i},\Over{i'}\in\ZZg$ 
a bilinear map
$\wedge:\GDiffpg{p}{\GMat{n}{m}}{i}\times\GDiffpg{p}{\GMat{n}{m}}{i'}\longrightarrow$\newline
$\GDiffpg{p+p'}{\GMat{n}{m}}{i+i'}$ by
\BGG
\label{d6}
\lp\omega\wedge\omega'\rp(D_{1},\cdots,D_{p+p'}) :=
\frac{1}{p!p'!}\sum_{\sigma\in\Perma{p+p'}}\Rom{sgn}\sigma\,
\gamma_{p+p'}(\sigma;\Over{D}_{1},\cdots,\Over{D}_{p+p'})\cdot
\qquad \qquad 
\\
\cdot(-1)^{\Over{i'}\sum_{l=1}^{p}\Over{D}_{\sigma(l)}}
\omega(D_{\sigma(1)},\cdots,D_{\sigma(p)})
\omega'(D_{\sigma(p+1)},\cdots,D_{\sigma(p+p')})
\nonumber
\EGG
for all homogeneous $D_{1},\cdots,D_{p+p'}\in\GDer{\GMat{n}{m}}$ and extend these by
bilinearity to $\GDiff{\GMat{n}{m}}$. With respect to it $\GDiff{\GMat{n}{m}}$ becomes a $\NNN\times\ZZg$-bigraded \CC-algebra.

Via
\BGG
\label{d7}
\lp L_{D_{0}}\omega\rp(D_{1},\cdots,D_{p}) &:=&
D_{0}\lp\omega(D_{1},\cdots,D_{p})\rp -
\\
& & -\sum_{l=1}^{p}(-1)^{\Over{D}_{0}(\Over{\omega}+\sum_{l'=1}^{l-1}\Over{D}_{l'})} 
\omega(D_{1},\cdots,\Comug{D_{0}}{D_{l}},\cdots,D_{p}) \, ,
\nonumber
\EGG
\BG
\label{d8}
\lp\imath_{D_{1}}\omega\rp(D_{2},\cdots,D_{p}) := \omega(D_{1},D_{2},\cdots,D_{p})
\qquad \qquad \qquad \qquad
\EG
and
\BGG
\label{d9}
d\omega(D_{0},\cdots,D_{p}) =
\sum_{l=0}^{p}
(-1)^{l+\Over{D}_{l}(\Over{\omega}+\sum_{l'=0}^{l-1}\Over{D}_{l'})}
L_{D_{l}}\lp\omega(D_{0},\cdots,\stackrel{\vee}{D_{l}},\cdots,D_{p})\rp + \qquad
\\
+\sum_{0\leq l<l'\leq p}
(-1)^{l'+\Over{D}_{l'}\sum_{l''=l+1}^{l'-1}\Over{D}_{l''}}
\omega(D_{0},\cdots,D_{l-1},\Comug{D_{l}}{D_{l'}},\cdots,\stackrel
{\vee}{D_{l'}},\cdots,D_{p})
\nonumber
\EGG
for all homogeneous $D_{0},D_{1},\cdots,D_{p}\in\GDer{\GMat{n}{m}}$ and all homogeneous
$\omega\in\GDiffp{p}{\GMat{n}{m}}$ ($\vee$ denotes omission), one defines homogeneous endomorphisms $\GDiff{\GMat{n}{m}}\longrightarrow\GDiff{\GMat{n}{m}}$ of bidegree
$(0,\Over{D}_{0})$, $(-1,\Over{D}_{0})$ and $(1,\Over{0})$ respectively. The assignments
$D\mapsto\imath_{D}$ and $D\mapsto L_{D}$ extend to \CC-linear maps
$\GDer{\GMat{n}{m}}\longrightarrow\End{\GDiff{\GMat{n}{m}}}$ and $L$ is a graded representation of $\GDer{\GMat{n}{m}}$ in particular. Furthermore the relations
\BGG
\label{d10}
d\circ d &=& 0 
\nonumber
\\
d\circ L_{D} &=& L_{D}\circ d 
\EGG
as well as
\BGG
\label{d11}
\imath_{D}\circ\imath_{D'} + (-1)^{\Over{D}\Over{D'}}\imath_{D'}\circ\imath_{D} &=& 0
\nonumber
\\
\lp L_{D}\circ\imath_{D'}-\imath_{D'}\circ L_{D}\rp\omega &=&
(-1)^{\Over{D}\Over{\omega}}\imath_{\Comug{D}{D'}}\omega
\\
\lp\imath_{D}\circ d+d\circ\imath_{D}\rp\omega &=& (-1)^{\Over{D}\Over{\omega}}L_{D}\omega 
\nonumber
\EGG
for all homogeneous $D,D'\in\GDer{\GMat{n}{m}}$ and all bihomogeneous
$\omega\in\GDiff{\GMat{n}{m}}$ are known from Lie superalgebra cohomology 
\cite{Scheunert8}.\newline
By analogy with the case of graded manifolds we call $d, L_{D}$ and $\imath_{D}$ exterior derivative, Lie derivative and interior product (with respect to a graded vector field
$D\in\GDer{\GMat{n}{m}}$). (\ref{d10}) and (\ref{d11}) tell us, that they fulfill exactly the
same relations as in the ``graded-commutative case'', but this observation remains also true for the graded wedge product (\ref{d6}) of graded forms.
\begin{proposition} 
\label{prop1d}
The relations
\BGG
\label{d12}
L_{D}\lp\omega\wedge\omega'\rp &=&
\lp L_{D}\omega\rp\wedge\omega' + (-1)^{\Over{D}\Over{\omega}}\omega\wedge L_{D}\omega'
\nonumber
\\
\imath_{D}\lp\omega\wedge\omega'\rp &=&
(-1)^{\Over{D}\Over{\omega'}}\lp\imath_{D}\omega\rp\wedge\omega' + 
(-1)^{p}\omega\wedge\imath_{D}\omega'
\\
d\lp\omega\wedge\omega'\rp &=&
\lp d\omega\rp\wedge\omega' + (-1)^{p}\omega\wedge d\omega'
\nonumber
\EGG
are fulfilled for all homogeneous $D\in\GDer{\GMat{n}{m}}$,
$\omega\in\GDiffp{p}{\GMat{n}{m}}$, $\omega'\in\GDiffp{p'}{\GMat{n}{m}}$. 
\end{proposition}
${\sl Proof}$: This can be shown exactly as in the case of graded manifolds. That is, one
starts with a direct proof of the second relation and proofs the other equations
inductively using the last two relations (\ref{d11}).
\hfill $\Box$ 

\vspace{2mm}
 
Because we interpret $d$ as exterior derivative, the Lie superalgebra cohomology of\newline
$\GDer{\GMat{n}{m}}$ with values in $\GMat{n}{m}$, 
\BG
\label{d13}
H(\GMat{n}{m}) \equiv \bigoplus_{p\in\NNNa}H^{p}(\GMat{n}{m}) :=
\frac{\Rom{ker}\,d}{\Rom{im}\,d} \, ,
\EG
has to be seen as analogue to the graded deRham-cohomology on graded manifolds. Via
\BG
\label{d14}
[\omega]\wedge[\omega'] := [\omega\wedge\omega']
\EG
the above graded derivation-based cohomology of $\GMat{n}{m}$ becomes a 
$\NNN\times\ZZg$-bigraded \CC-algebra and we will continue to study it in chapter 4.

\section{Homogeneous bases and the canonical graded 1-form}

Whereas the definitions and results of the preceding considerations apply to each 
$\ZZg$-graded, graded-central \CC-algebra we shall now employ more specific properties
of $\GMat{n}{m}$. There will result similar formulas as in ``ordinary'' matrix geometry
\cite{Djemai1,Dubois-Violette1,Madore10}, which is included as special case.

The sets $\GDiffpC{p}{\GMat{n}{m}}$ of graded $p$-forms with values in the graded center
of $\GMat{n}{m}$ form graded vector subspaces of $\GDiffp{p}{\GMat{n}{m}}$ for all
$p\in\NN$ and one can introduce
\BG
\label{b1}
\GDiffC{\GMat{n}{m}} := \bigoplus_{p\in\NNNa}\GDiffpC{p}{\GMat{n}{m}}
\EG
with $\GDiffpC{0}{\GMat{n}{m}}=\GCent{\GMat{n}{m}}$. $\GDiffC{\GMat{n}{m}}$ is a bigraded
subalgebra of $\GDiff{\GMat{n}{m}}$, whose product fulfills
\BG
\label{b2}
\omega\wedge\omega' = 
(-1)^{pp'+\Over{\omega}\Over{\omega}'}\omega'\wedge\omega
\EG
for all homogeneous $\omega\in\GDiffpC{p}{\GMat{n}{m}}, \omega'\in\GDiffpC{p'}{\GMat{n}{m}}$
and which is stable with respect to the whole Cartan calculus.\newline
Now let us introduce a homogeneous basis $\{\partial_{A}\}_{A=1,\cdots,n'+m'}$ of
$\GDer{\GMat{n}{m}}$ with $\partial_{1},\cdots,\partial_{n'}\in\GDerg{\GMat{n}{m}}{0}$,
$\partial_{n'+1},\cdots,\partial_{n'+m'}\in\GDerg{\GMat{n}{m}}{1}$, where we set
$n':=\Rom{dim}_{\CCa}\GDerg{\GMat{n}{m}}{0}$ and $m':=\Rom{dim}_{\CCa}\GDerg{\GMat{n}{m}}{1}$.
If $\{\eta^{A}\}_{A=1,\cdots,n'+m'}$ denotes the dual basis to 
$\{\partial_{A}\}_{A=1,\cdots,n'+m'}$ we can introduce a homogeneous basis
$\{\theta^{A}\}_{A=1,\cdots,n'+m'}$ of $\GDiffpC{1}{\GMat{n}{m}}$ by
\BG
\label{b3}
\theta^{A}(D) := \eta^{A}(D)1_{n+m}
\EG
for all $D\in\GDer{\GMat{n}{m}}$. Employing the standard isomorphisms between
graded-alternating maps and the graded exterior algebra \cite{Bartocci1,Scheunert8}
one deduces, that
\BG
\label{b4}
\lbr~\theta^{A_{1}}\wedge\cdots\wedge\theta^{A_{p}}~\Big
\vert~(A_{1},\cdots,A_{p})\in\GInd{n'}{m'}{p}~\rbr  
\EG
with
\BGG
\label{b5}
\GInd{n'}{m'}{p} := 
\lbr~(k_{1},\cdots,k_{p'},\alpha_{p'+1},\cdots,\alpha_{p})~\Big\vert~0\leq p'\leq p;
k_{1},\cdots,k_{p'}=1,\cdots,n'; 
\right.
\qquad \qquad \qquad
\\
\alpha_{p'+1},\cdots,\alpha_{p}=n'+1,\cdots,n'+m'; 
k_{1}<k_{2}<\cdots<k_{p'}<\alpha_{p'+1}\leq\cdots\leq\alpha_{p-1}\leq\alpha_{p}~\Big\}
\nonumber
\EGG
is a homogeneous basis of $\GDiffpC{p}{\GMat{n}{m}}, p\in\NN$.\newline
Because of 
\BG
\label{b6}
M\wedge\omega = (-1)^{\Over{M}\Over{\omega}}\omega\wedge M
\EG
for all homogeneous $M\in\GMat{n}{m}$ and all bihomogeneous $\omega\in\GDiffC{\GMat{n}{m}}$,
the $\NNN\times\ZZg$-bigraded \CC-algebras $\GDiff{\GMat{n}{m}}$ and
$\GMat{n}{m}\hat{\otimes}_{\CCa}\GDiffC{\GMat{n}{m}}$, where $\hat{\otimes}$ denotes
the tensor product of $\ZZg$-graded algebras, are canonically isomorphic. In particular we
can conclude:
\begin{proposition}
\label{prop1b}
The $\ZZg$-graded $\GMat{n}{m}$-bimodules $\GDiffp{p}{\GMat{n}{m}}$ are graded-free
for both multiplications and for all $p\in\NNN$. The set (\ref{b4}) determines a
homogeneous basis of the left (right), $\ZZg$-graded $\GMat{n}{m}$-module
$\GDiffp{p}{\GMat{n}{m}}$.\hfill $\Box$ 
\end{proposition}
Consequently every $\omega\in\GDiffp{p}{\GMat{n}{m}}$ can be written as
\BG
\label{b7}
\omega = \sum_{(A_{1},\cdots,A_{p})\in\GInda{n'}{m'}{p}}
\omega_{A_{1}\cdots A_{p}}\wedge\theta^{A_{1}}\wedge\cdots\wedge\theta^{A_{p}}
\EG
with unique coefficients $\omega_{A_{1}\cdots A_{p}}\in\GMat{n}{m}$. Explicitely these
coefficients are given by
\BG
\label{b8}
\omega_{A_{1}\cdots A_{p}} = 
(-1)^{\frac{1}{2}p''(p''-1)}\frac{1}{\prod_{l=1}^{n'+m'}N_{l}!}\,
\omega(\partial_{A_{1}},\cdots,\partial_{A_{p}}) \, ,
\EG
where $p''$ is the number of entries in $(A_{1},\cdots,A_{p})$
greater than $n'$ and $N_{l}$ is the number of entries 
in $(A_{1},\cdots,A_{p})$ being equal $l$.

In order to investigate graded derivations of $\GMat{n}{m}$ (we include the case
$n=m$ for the moment) let us denote by $\GGl{n}{m}$ the (complex) general linear Lie superalgebra and by $\GSl{n}{m}$ the (complex) special linear Lie superalgebra. The adjoint representation of $\GGl{n}{m}$ is at the same time a Lie superalgebra homomorphism
$\Rom{ad}:\GGl{n}{m}\longrightarrow\GDer{\GMat{n}{m}}$ and, as we will see, 
the structure of $\GDer{\GMat{n}{m}}$ and its Lie subsuperalgebras is determined by 
this homomorphism.
\begin{proposition}
\label{prop2b}
If $\Oger{L}$ is a Lie subsuperalgebra of $\GGl{n}{m}$ then
\BG
\label{b9}
\Oger{L}^{\Rom{ad}} := \Rom{im}\,\Rom{ad}\big\vert_{\Ogera{L}}
\EG
is a Lie subsuperalgebra of $\GDer{\GMat{n}{m}}$.
Contrary every Lie subsuperalgebra of \newline
$\GDer{\GMat{n}{m}}$ is of this form. There are two different cases:
\begin{enumerate}
\item[$\Rom{(i.)}$]
For $n\neq m$ the restriction of $\Rom{ad}$ to  
$\GSl{n}{m}$ is an Lie superalgebra isomorphism onto $\GDer{\GMat{n}{m}}$ and the 
various restrictions of $\Rom{ad}$ induce a bijective correspondence between
Lie subsuperalgebras of $\GSl{n}{m}$ and Lie subsuperalgebras of $\GDer{\GMat{n}{m}}$.
\item[$\Rom{(ii.)}$]
For $n=m$ there is no Lie subsuperalgebra $\Oger{L}$ of $\GGl{n}{n}$ such that the restriction of $\Rom{ad}$ to $\Oger{L}$ becomes an Lie superalgebra isomorphism onto $\GDer{\GMat{n}{n}}$.
\end{enumerate}
\end{proposition}
${\sl Proof}$: An even graded derivation of $\GMat{n}{m}$ is just an ordinary derivation 
of the $\CC$-algebra $\Mat{n+m}$ and these are inner, because the first Hochschild
cohomology group of $\Mat{n+m}$ with values in $\Mat{n+m}$ is trivial \cite{Pierce1}.
Introducing 
\BD
\Gamma := \lp
\begin{array}{ll}
1_{n} & 0 \\
0 & -1_{m}
\end{array}\rp 
\ED
we find for some $D\in\GDerg{\GMat{n}{m}}{1}$ and all homogeneous $M\in\GMat{n}{m}$
\BD
DM = \Rom{ad}\lp\frac{1}{2}(D\Gamma)\Gamma\rp(M) \, ,
\ED
from which we can conclude, that $D$ is inner. Consequently, if $\Oger{D}$ is a
Lie subsuperalgebra of $\GDer{\GMat{n}{m}}$, then $\Oger{L}:=\Rom{ad}^{-1}(\Oger{D})$
is a Lie subsuperalgebra of $\GGl{n}{m}$ with $\Oger{L}^{\Rom{ad}}=\Oger{D}$.
(i.) and (ii.) are consequences of $1_{n+m}\not\in\GSl{n}{m}$ for $n\neq m$ respectively
$1_{2n}\in\Comug{\GGlg{n}{n}{1}}{\GGlg{n}{n}{1}}$.
\hfill $\Box$ 

\vspace{2mm}

The ultimate reason for restricting our geometric investigation to the case $n\neq m$ lies
in the existence of the Lie superalgebra isomorphism
$\Rom{ad}:\GSl{n}{m}\longrightarrow\GDer{\GMat{n}{m}}$. The elements of every homogeneous basis $\{\partial_{A}\}_{A=1,\cdots,n'+m'}$ of $\GDer{\GMat{n}{m}}$ are given by
\BG
\label{b10}
\partial_{A} = \Rom{ad}\,E_{A} \, ,
\EG
where $\{ E_{A}\}_{A=1,\cdots,n'+m'}$ is a homogeneous basis of $\GSl{n}{m}$ and we have
$n'=n^{2}+m^{2}-1, m'=2nm$ in particular. Moreover, the structure constants $c_{AB}^{C}$
appearing in
\BG
\label{b11}
\Comug{\partial_{A}}{\partial_{B}} = 
\sum_{C=1}^{(n+m)^{2}-1}c_{AB}^{C}\partial_{C}
\EG
are the structure constants of the homogeneous $\GSl{n}{m}$-basis 
$\{ E_{A}\}_{A=1,\cdots,(n+m)^{2}-1}$ and one deduces the nice formulas
\BG
\label{b12}
dE_{A} = -\sum_{B,C=1}^{(n+m)^{2}-1}c_{AB}^{C}E_{C}\wedge\theta^{B} 
\EG
and
\BG
\label{b13}
d\theta^{A} = 
\frac{1}{2}\sum_{B,C=1}^{(n+m)^{2}-1}c_{BC}^{A}\theta^{C}\wedge\theta^{B} \, .
\EG
The even graded $1$-form
\BG
\label{b14}
\Theta := 
\sum_{A=1}^{(n+m)^{2}-1}E_{A}\wedge\theta^{A} 
\EG
will be called canonical graded 1-form, because it plays a distinguished role.
\begin{proposition}
\label{prop3b}
The definition of $\Theta$ is independent of the choice of the 
homogeneous basis of $\GDer{\GMat{n}{m}}$. $\Theta$ is ($\GDer{\GMat{n}{m}}$-)invariant and 
this property determines $\Theta$ up to constant multiples.
Furthermore its exterior differential fulfills
\BG
\label{b15}
d\Theta = \Theta\wedge\Theta 
\EG
and the exterior differential of each $M\in\GMat{n}{m}$ can be expressed according to
\BG
\label{b16}
dM = \Comug{\Theta}{M} \equiv \Theta\wedge M - M\wedge\Theta \, .
\EG
\end{proposition}
${\sl Proof}:$ Beside the uniqueness statement only simple calculations are involved
(for which one can use (\ref{b12}) and (\ref{b13}) advantageously). 
The irreducibility of the adjoint representation of $\GSl{n}{m}$ \cite{Cornwell1}
guarantees, that 
\BD
L_{D}\omega = 0 \, , \quad \omega\in\GDiffp{1}{\GMat{n}{m}} \, ,
\ED
for all $D\in\GDer{\GMat{n}{m}}$ implies $\omega=c\Theta, c\in\CC$. \hfill $\Box$ 

\vspace{2mm}

Finally we note, that $\GDiff{\GMat{n}{m}}$ is in a certain sense minimal.
\begin{proposition}
\label{prop4b}
(\ref{b12}) can be inverted according to
\BG
\label{b17}
\theta^{A} = 
4(n-m)^{2}\sum_{B,C,D=1}^{(n+m)^{2}-1}(-1)^{\Over{E}_{B}\Over{E}_{D}}
K^{AB}K^{CD}E_{C}E_{B}\wedge dE_{D} \, ,
\EG
where $K$ is the Killing form of $\GSl{n}{m}$ and $K^{AB}$ denote the components of the inverse matrix of $(K(E_{A},E_{B}))$. Consequently, if $\Omega$ is differential subalgebra 
of $\GDiff{\GMat{n}{m}}$ containing $\GMat{n}{m}$, then $\Omega=\GDiff{\GMat{n}{m}}$. 
\end{proposition}
${\sl Proof}:$ The minimality statement follows from (\ref{b17}) because of proposition
\ref{prop1b}. In order to show (\ref{b17}) one uses (\ref{b12}) and expands the
threefold product of the basis elements $E_{A}$ according to (\ref{a4}). Using the
results of proposition \ref{prop1a} (\ref{b17}) follows.
\hfill $\Box$ 

\vspace{2mm}

The second part of proposition \ref{prop4b} can be stated differently: The canonical even
algebra homomorphisms from the (intrinsic) $\ZZg$-graded universal differential envelope 
of $\GMat{n}{m}$ to $\GDiff{\GMat{n}{m}}$ (see \cite{Coquereaux1,Kastler1} for a precise definition) is onto. The restriction of this homomorphism to the corresponding first-order
differential calculi is an isomorphism.

\section{Cohomology and the noncommutative body map}

We will call the even, surjective $\CC$-linear map 
\BG
\label{c1}
\beta:\GMat{n}{m}\longrightarrow\Mat{\Under{n}}
\qquad \mbox{with} \quad
\Under{n} := \lbr
\begin{array}{ll}
n \;\; & \mbox{if}~n>m \\
m \;\; & \mbox{if}~n<m 
\end{array}
\right.
\EG
defined by
\BG
\label{c2}
\beta(M) \equiv
\beta\lp\lp
\begin{array}{ll}
M_{1} & M_{2} \\
M_{3} & M_{4} 
\end{array}
\rp\rp :=
\lbr
\begin{array}{ll}
M_{1} \;\; & \mbox{if}~\Under{n}=n \\
M_{4} \;\; & \mbox{if}~\Under{n}=m
\end{array}
\right. 
\EG
canonical body map of $\GMat{n}{m}$. A justification for choosing this terminology
will result from the investigation of its properties: They are completely analogous to the corresponding map of graded manifolds if one takes the noncommutativity of $\GMat{n}{m}$ 
and its ``body'' $\Mat{\Under{n}}$ appropriately into account. In order to
distinguish between ``objects'' on $\GMat{n}{m}$ and corresponding ``objects'' on the body
we underline the latter.

The restriction of $\beta$ to $\GCent{\GMat{n}{m}}$ is an even algebra homomorphism onto $\Cent{\Mat{\Under{n}}}$ and by 
\BG
\label{c3}
\iota(\Under{M}) := \lbr
\begin{array}{ll}
\lp
\begin{array}{cc}
\Under{M} & 0 \\
0 & \frac{\DS 1}{\DS n}\Rom{Tr}(\Under{M})1_{m}
\end{array}
\rp \;\; & \mbox{if}~\Under{n}=n \\
\\
\lp
\begin{array}{cc}
\frac{\DS 1}{\DS m}\Rom{Tr}(\Under{M})1_{n} & 0 \\
0 & \Under{M} 
\end{array}
\rp \;\; & \mbox{if}~\Under{n}=m
\end{array}
\right. \, ,
\EG
we can introduce an even, injective $\CC$-linear map
$\iota:\Mat{\Under{n}}\longrightarrow\GMat{n}{m}$, which is right-inverse to $\beta$ on the
one hand and whose restriction to $\Cent{\Mat{\Under{n}}}$ is an even algebra homomorphism
into $\GCent{\GMat{n}{m}}$ on the other hand.\newline
Analogous to the body map of graded manifolds $\beta$ induces a Lie algebra homomorphism $\hat{\beta}:\GDerg{\GMat{n}{m}}{0}\longrightarrow\Der{\Mat{\Under{n}}}$ via
\BG
\label{c5}
\hat{\beta}(D)\beta(M) := \beta(DM) 
\EG
for all $M\in\GMat{n}{m}$. $\hat{\beta}$ is surjective because of
\BG
\label{c6}
\hat{\beta}(\Rom{ad}\,E) = \Rom{ad}\,\beta(E)
\EG
for all $E\in\GSlg{n}{m}{0}$ and in addition
$\hat{\iota}:\Der{\Mat{\Under{n}}}\longrightarrow\GDerg{\GMat{n}{m}}{0}$,
\BG
\label{c7}
\hat{\iota}(\Rom{ad}\,\Under{E}) := \Rom{ad}\,\iota(\Under{E})
\EG
is a Lie algebra homomorphism right-inverse to $\hat{\beta}$. \newline
Now we can introduce even \CC-linear maps $\beta^{(p)}:\GDiffp{p}
{\GMat{n}{m}}\longrightarrow\Diffp{p}{\Mat{\Under{n}}}, p\in\NN$, by
\BG
\label{c8}
\lp\beta^{(p)}(\omega)\rp\lp\Under{D}_{1},\cdots,\Under{D}_{p}\rp :=
\beta\lp\omega\lp\hat{\iota}(\Under{D}_{1}),\cdots,\hat{\iota}(\Under{D}_{p})\rp\rp
\EG
for all $\Under{D}_{1},\cdots,\Under{D}_{p}\in\Der{\Mat{\Under{n}}}$.
$\GSlg{n}{m}{0}$ is canonically isomorphic to $\Sl{n}\oplus\Gl{1}\oplus\Sl{m}$ and we
can choose a homogeneous basis $\{ E_{A}\}_{A=1,\cdots,(n+m)^{2}-1}$ of $\GSl{n}{m}$
such that $E_{1},\cdots,E_{\Under{n}^{2}-1}$ lie in the isomorphic copy of $\Sl{\Under{n}}$
and $E_{\Under{n}^{2}},\cdots,E_{n^{2}+m^{2}-1}$ in the isomorphic copy of 
$\Gl{1}\oplus\Sl{\Under{m}}$ with $\Under{m}:=\mbox{min}\{ n,m\}$. Then the elements $\beta(E_{k}):=\Under{E}_{k}, k=1,\cdots,\Under{n}^{2}-1$, form a basis of $\Sl{\Under{n}}$.
Denoting the elements of the basis of $\Diffp{1}{\Mat{\Under{n}}}$ corresponding with
$\{ \Under{E}_{k}\}_{k=1,\cdots,\Under{n}^{2}-1}$ according to (\ref{b3}) and (\ref{b10})
with $\Under{\theta}^{k}$ the action of the maps $\beta^{(p)}$ can be described 
alternatively by
\BGG
\label{c9}
\beta^{(p)}(\omega) &\equiv&
\beta^{(p)}\lp\sum_{(A_{1},\cdots,A_{p})\in\GInda{n^{2}+m^{2}-1}{2nm}{p}}
\omega_{A_{1}\cdots A_{p}}\wedge\theta^{A_{1}}\wedge\cdots\wedge\theta^{A_{p}}\rp =
\nonumber
\\
&=&
\sum_{(k_{1},\cdots,k_{p})\in\GInda{\Under{n}^{2}-1}{0}{p}}
\beta\lp\omega_{k_{1}\cdots k_{p}}
\rp\wedge\Under{\theta}^{k_{1}}\wedge\cdots\wedge\Under{\theta}^{k_{p}} \, .
\EGG
If we set $\beta^{(0)}\equiv\beta$ the maps $\beta^{(p)}, p\in\NNN$, extend uniquely to a 
bihomogeneous, \CC-linear map $\GDiff{\GMat{n}{m}}\longrightarrow\Diff{\Mat{\Under{n}}}$
of bidegree $(0,\Over{0})$ , which we again denote by $\beta$. Because of (\ref{c9}) 
$\beta$ is onto and its restriction to $\GDiffC{\GMat{n}{m}}$ is an surjective homomorphism
of bigraded \CC-algebras onto $\DiffC{\Mat{\Under{n}}}$. Furthermore $\beta$ fulfills
\BG
\label{c10}
\beta\circ L_{D} = L_{\hat{\beta}(D)}\circ\beta
\EG
for all $D\in\GDerg{\GMat{n}{m}}{0}$ as well as
\BG
\label{c11}
\beta\circ d = d\circ\beta \, .
\EG
Consequently $\beta$ induces a homomorphism 
$H(\beta):H(\GMat{n}{m})\longrightarrow H(\Mat{\Under{n}})$ of cohomologies in the usual way. Analogous to graded manifold theory \cite{Kostant1} this map is an isomorphism.  
\begin{proposition}
\label{prop1c}
$H(\beta)$ is an isomorphism of bigraded \CC-algebras, such that both cohomologies
$H(\GMat{n}{m})$ and $H(\Mat{\Under{n}})$ are isomorphic to the Lie algebra cohomology
$H(\Sl{\Under{n}};\CC)$ of $\Sl{\Under{n}}$ with trivial coefficients.
\end{proposition}
${\sl Proof}$: Using the results of \cite{Scheunert8} as well as 
$\GDer{\GMat{n}{m}}=\GSl{n}{m}^{\Rom{ad}}$ we find the sequence 
\BA
H(\GMat{n}{m}) &\cong & H(\GSl{n}{m};\GMat{n}{m}) \cong
H(\GSl{n}{m};\CC\, 1_{n+m})\oplus H(\GSl{n}{m};\GSl{n}{m}) \cong
\\
&\cong & H(\GSl{n}{m};\CC\, 1_{n+m}) \cong
H(\GSl{n}{m};\CC)
\EA
of natural isomorphisms between Lie superalgebra cohomologies. In particular we
have $H(\Mat{\Under{n}})\cong H(\Sl{\Under{n}};\CC)$ (as $\NNN$-graded \CC-algebra), which is well-known from matrix geometry \cite{Djemai1,Dubois-Violette1,Dubois-Violette5}.
Combining the above result with the calculations of the cohomology of $\GSl{n}{m}$
with trivial coefficients \cite{Fuks1,Fuks2} one can conclude, that $H(\beta)$ is
an isomorphism of bigraded \CC-algebras.
\hfill $\Box$

\section{Noncommutative graded symplectic geometry}

Generalizing the situation on graded manifolds \cite{Kostant1} we call an even, closed
graded 2-form $\omega\in\GDiffp{2}{\GMat{n}{m}}$ graded symplectic structure on $\GMat{n}{m}$,
if the equation
\BG
\label{sy1}
\omega(D,D_{M}) = DM
\EG
for all $D\in\GDer{\GMat{n}{m}}$ possesses a unique solution $D_{M}\in\GDer{\GMat{n}{m}}$
for each $M\in\GMat{n}{m}$. The graded vector fields $D_{M}\in\GDer{\GMat{n}{m}}$
are called Hamiltonian and the set of all graded Hamiltonian vector fields is denoted
by $\GHam{\omega}$.\newline
If $\omega\in\GDiffp{2}{\GMat{n}{m}}$ is graded symplectic structure on $\GMat{n}{m}$ the
assignment $M\mapsto D_{M}$ defines an even $\CC$-linear map 
$D^{\omega}:\GMat{n}{m}\longrightarrow\GHam{\omega}\subseteq\GDer{\GMat{n}{m}}$ and
one can conclude, that (\ref{sy1}) is equivalent to
\BG
\label{sy2}
\imath_{D_{M}}\omega + dM = 0 \, .
\EG
Using (\ref{d11}) we find
\BG
\label{sy3}
L_{D_{M}}\omega = 0 
\EG
for all $D_{M}\in\GHam{\omega}$, that is a graded symplectic structure on $\GMat{n}{m}$ is
- as usual - invariant with respect to all graded Hamiltonian vector fields.\newline
Via 
\BG
\label{sy4}
\Poissong{M}{M'}{\omega} := \omega(D_{M},D_{M'})
\EG
for all $M,M'\in\GMat{n}{m}$ we can introduce a graded Poisson bracket, which has the
analogous properties as its graded-commutative pendant.
\begin{proposition}
\label{prop1sy} 
$(\GMat{n}{m},\Poissong{\cdot}{\cdot}{\omega})$ is a $\CC$-Lie superalgebra and the
graded Poisson bracket fulfills in addition
\BGG
\label{sy5}
\Poissong{M}{M'M''}{\omega} &=&
\Poissong{M}{M'}{\omega}M'' +
(-1)^{\Over{M}\Over{M}'}M'\Poissong{M}{M''}{\omega}
\nonumber
\\
\Poissong{1_{n+m}}{M}{\omega} &=& 0
\EGG
for all homogeneous $M,M',M''\in\GMat{n}{m}$. Moreover, the map
$D^{\omega}:\GMat{n}{m}\longrightarrow\GHam{\omega}$ is a homomorphism of Lie superalgebras and 
\BG
\label{sy6}
\GHam{\omega}=\GDer{\GMat{n}{m}} \, .
\EG
\end{proposition}
${\sl Proof}$: The properties of $\Poissong{\cdot}{\cdot}{\omega}$ and of $D^{\omega}$ 
result from the defining properties of the graded symplectic structure $\omega$.
From the irreducibility of the adjoint representation of $\GSl{n}{m}$ one can deduce
$\mbox{ker}\,D^{\omega}=\CC\,1_{n+m}$ on the one hand and the injectivity of
$D^{\omega}\vert_{\GSla{n}{m}}$ on the other hand. Then (\ref{sy6}) follows because of
$\GDer{\GMat{n}{m}}=\GSl{n}{m}^{\Rom{ad}}$.
\hfill $\Box$ 

\vspace{2mm}

There exists an essentially unique graded symplectic structure on $\GMat{n}{m}$.
\begin{proposition}
\label{prop2sy}
$d\Theta$ is a graded symplectic structure on $\GMat{n}{m}$ and up to complex multiples
it is the only one. The corresponding graded Poisson bracket is given by
\BG
\label{sy7}
\Poissong{M}{M'}{d\Theta} = \Comug{M}{M'}
\EG
for all $M,M'\in\GMat{n}{m}$.
\end{proposition}
${\sl Proof}$: The exact, even graded $2$-form $c\,d\Theta, c\in\CC\setminus\{ 0\}$
induces via (\ref{sy1}) a homomorphism\newline
$D^{c\,d\Theta}:\GMat{n}{m}\longrightarrow\GDer{\GMat{n}{m}}$,
\BG
\label{sy8}
D^{c\,d\Theta}(M) = \frac{1}{c}\Rom{ad}M
\EG
of Lie superalgebras and the corresponding graded Poisson bracket is given by
$\Poissong{M}{M'}{c\,d\Theta} = c^{-1}\Comug{M}{M'}$.
The uniqueness property is a consequence of proposition \ref{prop2b}, (\ref{sy6}) 
and Schur's Lemma. \hfill $\Box$ 

\vspace{2mm}

Consequently the extension of the body map $\beta$ maps a graded symplectic structure 
$\omega$ onto a symplectic structure $\beta(\omega)$. Moreover one has
\BG
\label{sy9}
\hat{\beta}\lp D^{\omega}(M)\rp = D^{\beta(\omega)}\lp\beta(M)\rp
\EG
for all even graded (Hamiltonian) vector fields as well as
\BG
\label{sy10}
\beta\lp\Poissong{M}{M'}{\omega}\rp =
\Poisson{\beta(M)}{\beta(M')}{\beta(\omega)}
\EG
for the graded Poisson bracket of $M,M'\in\GMatg{n}{m}{0}$. That is, the relation between
$\GMat{n}{m}$ and its body is analogous to the one for graded symplectic manifolds and their
respective underlying manifolds.

\section{Graded vector bundles over graded matrix algebras}

As a synthesis of the definition of graded vector bundles over graded manifolds
\cite{Bartocci1,Hernandez1,Kostant1} and the idea how to introduce vector bundles
in noncommutative geometry \cite{Connes1,Gracia-Bondia1} we interpret left, $\ZZg$-graded,
finitely generated (graded-projective) $\GMat{n}{m}$-modules as graded vector bundles over
$\GMat{n}{m}$ and even $\GMat{n}{m}$-module homomorphisms between such modules as graded
vector bundle homomorphisms. We note, that the specifying property of graded projectivity 
is redundant in the context of left, $\ZZg$-graded $\GMat{n}{m}$-modules,
because on the one hand graded-projective means $\ZZg$-graded plus projective
\cite{Nastasescu1} and on the other hand every left $\Mat{n+m}$-module is projective
\cite{Pierce1}.\newline
Let us denote by $\GMatnm{n}{m}{r}{s}, r,s\in\NNN, r+s\in\NN,$ the $\CC$-vector space $\Matnm{n+m}{r+s}$ together with the $\ZZg$-grading defined by
\BGG
\label{v1}
\GMatnmg{n}{m}{r}{s}{0} &:=&
\lbr~v=\lp
\begin{array}{ll}
v_{1} & 0 \\
0 & v_{4} 
\end{array}\rp
\Big\vert~v_{1}\in\Matnm{n}{r}, 
v_{4}\in\Matnm{m}{s}~\rbr
\nonumber
\\
\GMatnmg{n}{m}{r}{s}{1} &:=&
\lbr~v=\lp
\begin{array}{ll}
0 & v_{2} \\
v_{3} & 0 
\end{array}\rp
\Big\vert~v_{2}\in\Matnm{n}{s}, 
v_{3}\in\Matnm{m}{r}~\rbr \, .
\EGG 
With respect to ordinary matrix multiplication $\GMatnm{n}{m}{r}{s}$ becomes a left,
$\ZZg$-graded, finitely generated $\GMat{n}{m}$-module and these examples constitute
essentially all graded vector bundles over $\GMat{n}{m}$. 
\begin{proposition}
\label{prop1v} 
If $\Cal{V}$ is a graded vector bundle over $\GMat{n}{m}$ then there exist unique
numbers $r,s\in\NNN, r+s\in\NN$ and a graded vector bundle isomorphism
$\phi:\Cal{V}\longrightarrow\GMatnm{n}{m}{r}{s}$. $\Cal{V}$ is graded-free if and only if
there are natural numbers $p,q\in\NNN, p+q\in\NN,$ such that
\BGG
\label{v2}
pn + qm &=& r
\nonumber
\\
pm + qn &=& s \, .
\EGG
\end{proposition}
${\sl Proof}$: The existence of the isomorphisms are implied by the graded simplicity of
$\GMatnm{n}{m}{1}{0}$ and $\GMatnm{n}{m}{0}{1}$ and the fact, that every 
left, $\ZZg$-graded, finitely generated $\GMat{n}{m}$-module is the homomorphic image
of a left, $\ZZg$-graded, graded-free $\GMat{n}{m}$-module with homogeneous basis of suitable cardinality $p\vert q$. Because all $\GMat{n}{m}$-module isomorphisms are $\CC$-vector
space isomorphisms in particular, the uniqueness statement and (\ref{v2}) follow.
\hfill $\Box$

\vspace{2mm}

After this ``miniature-classification'' we develop graded differential geometry on a 
fixed graded vector bundle $\Cal{V}$ generalizing the treatment of noncommutative geometry 
\cite{Connes1,Djemai1,Gracia-Bondia1,Madore10} on the one hand and
the one of supergeometry \cite{Bartocci1} on the other hand.\newline
So we first define the set $\GDiff{\Cal{V}}$ of $\Cal{V}$-valued graded forms according to 
\BG
\label{v3}
\GDiff{\Cal{V}} \equiv \bigoplus_{p\in\NNNa}\GDiffp{p}{\Cal{V}} :=
\GDiff{\GMat{n}{m}}\hat{\otimes}_{\GMata{n}{m}}\Cal{V} \, .
\EG
$\GDiff{\Cal{V}}$ is a left $\NNN\times\ZZg$-bigraded $\GDiff{\GMat{n}{m}}$-module
in a natural way and each $\GDiffp{p}{\Cal{V}}, p\in\NNN$, is a left, $\ZZg$-graded,
finitely generated $\GMat{n}{m}$-module in particular. The product will again be denoted
by $\wedge$.\newline
A connection on $\Cal{V}$ is an even $\CC$-linear map 
$\nabla:\Cal{V}\longrightarrow\GDiffp{1}{\Cal{V}}$ such that
\BG
\label{v4}
\nabla\lp Mv\rp = dM\otimes v + M\wedge\nabla v
\EG
is fulfilled for all $M\in\GMat{n}{m}, v\in\Cal{V}$. Connections always exist due to (graded)
projectivity.
\begin{proposition}
\label{prop2v} 
Let $\Cal{V}$ be a graded vector bundle over $\GMat{n}{m}$. Then there exists a  
graded-free vector bundle $\Cal{V}^{p\vert q}$ over $\GMat{n}{m}$ with homogeneous basis 
$\{~\varepsilon_{A}~\vert~\varepsilon_{A}\in\Cal{V}^{p\vert q}_{\Over{0}}, A=1,\cdots,p;
\varepsilon_{A}\in\Cal{V}^{p\vert q}_{\Over{1}}, A=p+1,\cdots,p+q \}, p,q\in\NNN, p+q\in\NN$,
together with an even, idempotent endomorphism 
$P:\Cal{V}^{p\vert q}\longrightarrow\Cal{V}^{p\vert q}$ and an isomorphism $\varphi:\Cal{V}\longrightarrow\Rom{im}P$ of $\ZZg$-graded $\GMat{n}{m}$-modules. The map
$\nabla_{d}:\Cal{V}^{p\vert q}\longrightarrow\GDiffp{1}{\Cal{V}^{p\vert q}}$ defined by
\BG
\label{v5}
\nabla_{d}(v) \equiv \nabla_{d}\lp\sum_{A=1}^{p+q}v^{A}\varepsilon_{A}\rp :=
\sum_{A=1}^{p+q}dv^{A}\otimes\varepsilon_{A}
\EG
is a connection on $\Cal{V}^{p\vert q}$ and
\BG
\label{v6}
\nabla_{Pd} := 
\Rom{Id}_{\GDiffp{1}{\GMata{n}{m}}}\otimes\varphi^{-1}\circ 
\Rom{Id}_{\GDiffp{1}{\GMata{n}{m}}}\otimes P\circ\nabla_{d}\circ\varphi
\EG
is a connection on $\Cal{V}$. A map $\nabla:\Cal{V}\longrightarrow\GDiffp{1}{\Cal{V}}$ is 
a connection on $\Cal{V}$ if and only if it is of the form
\BG
\label{v7}
\nabla = \nabla_{Pd} + \alpha \, ,
\EG
where $\alpha:\Cal{V}\longrightarrow\GDiffp{1}{\Cal{V}}$
is an even homomorphism of $\ZZg$-graded $\GMat{n}{m}$-modules.
\end{proposition}
${\sl Proof}$: Analogous to the ungraded case \cite{Connes1,Gracia-Bondia1}.
\hfill $\Box$ 

\vspace{2mm}

Quite generally we will denote the $\ZZg$-graded $\GMat{n}{m}$-bimodule of graded 
$(p+q)\times(p+q)$-matrices over a $\ZZg$-graded bimodule $\Cal{B}$ with
$\GMatk{p}{q}{\Cal{B}}$. It is a $\GMat{n}{m}$-bimodule in a natural way and
$\ZZg$-graded by declaring those matrices with even diagonal entries and odd off-diagonal entries as even and those with odd diagonal entries and even off-diagonal entries as odd. 
Adopting the notation of the above proposition we introduce homogeneous generators
\BG
\label{v9}
\eta_{A} := \varphi^{-1}\circ P(\varepsilon_{A})
\EG
of $\Cal{V}$ as well as an even matrix $(P^{B}_{A})\in\GMatkg{p}{q}{\GMat{n}{m}}{0}$
via
\BG
\label{v10}
P(\varepsilon_{A}) =: \sum_{B=1}^{p+q}P_{A}^{B}\varepsilon_{B} \, .
\EG
Then
\BG
\label{v11}
\lp\nabla-\nabla_{Pd}\rp(\eta_{A}) = \alpha(\eta_{A}) =:
\sum_{B=1}^{p+q}\alpha_{A}^{B}\otimes\eta_{B}
\EG

establishes a bijective correspondence between the set of all connections on $\Cal{V}$ and
the set $P\GMatkg{p}{q}{\GDiffp{1}{\GMat{n}{m}}}{0}P$, which consists of those
$(\alpha_{A}^{B})\in\GMatkg{p}{q}{\GDiffp{1}{\GMat{n}{m}}}{0}$ fulfilling
$\alpha_{A}^{B}=\sum_{C,D=1}^{p+q}P_{A}^{C}\wedge\alpha_{C}^{D}\wedge P_{D}^{B}$
(for $\Cal{V}=\Cal{V}^{p\vert q}$ set $P=\varphi=\Rom{Id}_{\Cal{V}^{p\vert q}}$).
The graded 1-forms $\alpha_{A}^{B}$ are called connection forms of the 
connection $\nabla=\nabla_{Pd}+\alpha$.\newline
If $\nabla$ is a connection on a graded vector bundle $\Cal{V}$ we can introduce
a $\CC$-linear map $\GDiff{\Cal{V}}\longrightarrow\GDiff{\Cal{V}}$, again denoted by
$\nabla$, via
\BG
\label{v12}
\nabla\lp\omega\otimes v\rp = d\omega\otimes v + (-1)^{p}\omega\wedge\nabla v
\EG
for all $v\in\Cal{V}, \omega\in\GDiffp{p}{\GMat{n}{m}}, p\in\NNN$. This 
homogeneous map of bidegree $(1,\Over{0})$ extends the original connection
if we identify $\Cal{V}$ with $\GDiffp{0}{\Cal{V}}$. Moreover it fulfills
\BG
\label{v13}
\nabla\lp\omega'\wedge\omega\otimes v\rp = 
d\omega'\wedge(\omega\otimes v) + (-1)^{p'}\omega'\wedge\nabla(\omega\otimes v)
\EG
for all $v\in\Cal{V}, \omega\in\GDiffp{p}{\GMat{n}{m}}, \omega'\in\GDiffp{p'}{\GMat{n}{m}}, p,p'\in\NNN$, and this property determines the extension of the connection 
uniquely.\newline
The curvature of a connection $\nabla$ on a graded vector bundle $\Cal{V}$ is defined as
\BG
\label{v14}
\nabla^{2} \equiv \nabla\circ\nabla: \Cal{V}\longrightarrow\GDiffp{2}{\Cal{V}} \, .
\EG
It is an even homomorphism of $\ZZg$-graded $\GMat{n}{m}$-modules and one can
describe its action on an element $v=\sum_{A=1}^{p+q}\varphi(v)^{A}\eta_{A}\in\Cal{V}$
according to
\BG
\label{v15}
\nabla^{2}(v) =: \sum_{A,B=1}^{p+q}\varphi(v)^{A}\wedge R^{B}_{A}\otimes \eta_{B}
\EG
with a uniquely determined matrix
$(R^{B}_{A})\in P\GMatkg{p}{q}{\GDiffp{2}{\GMat{n}{m}}}{0}P$.  The graded $2$-forms
$R^{B}_{A}$ are called curvature forms and they can be expressed according to
\BG
\label{v16}
R^{B}_{A} = -\sum_{C=1}^{p+q}\alpha_{A}^{C}\wedge\alpha_{C}^{B} +
\sum_{C,D=1}^{p+q}\lp
P_{A}^{C}\wedge d\alpha_{C}^{D}\wedge P_{D}^{B}-P_{A}^{C}\wedge dP_{C}^{D}\wedge dP_{D}^{B}\rp
\EG
in terms of the connection forms $\alpha_{A}^{B}$ of the connection. 
Moreover they have to fulfill the Bi\-an\-chi identity
\BG
\label{v17}
\sum_{C,D=1}^{p+q}P_{A}^{C}\wedge dR_{C}^{D}\wedge P_{D}^{B} -
\sum_{C=1}^{p+q}\lp
\alpha_{A}^{C}\wedge R_{C}^{B}-R_{A}^{C}\wedge\alpha_{C}^{B}\rp = 0 \, .
\EG

Let us finally analyze the space of flat connections, that is the set of all
connections with vanishing curvature. We will not do this in complete generality
but only for a graded-free vector bundle $\Cal{V}^{1\vert 0}$ with an even 
basis element $\varepsilon$.
\begin{proposition}
\label{prop3v} 
A connection on $\Cal{V}^{1\vert 0}$ is flat if and only if its connection form \newline
$\alpha\in\GDiffpg{1}{\GMat{n}{m}}{0}$ is either given by
\BG
\label{v18}
\alpha = \Theta
\EG
or by 
\BG
\label{v19}
\alpha = \Theta - \sum_{A=1}^{(n+m)^{2}-1}f(E_{A})\wedge\theta^{A} \, ,
\EG
where $\{ E_{A}\}$ is the homogeneous basis of $\GSl{n}{m}$ ``corresponding'' with
$\{\theta^{A}\}$ and $f$ is some automorphism of $\GSl{n}{m}$.
\end{proposition}
${\sl Proof}$: Let us introduce an even graded 1-form 
$\rho=\sum_{A=1}^{(n+m)^{2}-1}\rho_{A}\wedge\theta^{A}$ according to 
$\alpha=:\Theta-\rho$. Using proposition \ref{prop3b} we find, that the curvature form  
is given by
\BG
\label{v20}
R = \frac{1}{2}\sum_{A,B=1}^{(n+m)^{2}-1}\Omega_{AB}\wedge\theta^{A}\wedge\theta^{B}
\EG
with
\BG
\label{v21}
\Omega_{AB} = \Comug{\rho_{B}}{\rho_{A}} - \sum_{C=1}^{(n+m)^{2}-1}c_{BA}^{C}\rho_{C} \, .
\EG
Because the vanishing of the curvature is equivalent to $\Omega_{AB}=0$ the proposition
follows from the simplicity of $\GSl{n}{m}$.
\hfill $\Box$ 

\vspace{2mm}

That is, we have the same situation as in ordinary matrix geometry 
\cite{Djemai1,Dubois-Violette1}: There exist different ``classes'' of flat connections.
Here ``class'' refers to the action of the group of automorphisms of the graded vector
bundle on the space of connections, which can be introduced analogous to the ungraded
case. The connection $\nabla_{d}$ and the one associated with the connection form $\Theta$ will lie in different classes, because the latter is invariant. However, if one does not
restrict the space of connections by a suitable compatibility requirement with respect
to a graded hermitian structure there will exist even more than two classes of flat
connections.

\section{Concluding remarks}

We have developed the graded differential geometry of graded matrix algebras and shown
that the results of matrix geometry \cite{Djemai1,Dubois-Violette1} carry over to the
$\ZZg$-graded setting. In addition we found a natural noncommutative analogue of the 
body map, which allows us to view graded matrix geometries as true noncommutative
generalizations of graded manifolds.\newline
Whereas in ordinary differential geometry one integrates forms this is not true in
supergeometry. Except from the before mentioned body map, which plays a central role in the
global theory of Berezin integration \cite{Hernandez2}, we completely excluded the integral geometry of graded matrix algebras. We plan to treat this together with metric aspects in a separate work.\newline
Beside its immediate application for the construction of (graded) differential calculi
on fuzzy (super)manifolds \cite{Madore10,Grosse15} the developments of this
article offer another perspective.
The extension of space-time by matrix geometries led to interesting new gauge models. In 
particular the existence of different gauge orbits of flat connections in matrix geometry 
is the origin of the appearance of the Higgs effect \cite{Djemai1,Dubois-Violette10}. 
The possibility of extending the structures of matrix geometry to $\ZZg$-graded matrix
algebras suggests to think about similar ``supersymmetric'' noncommutative extensions of 
space-time. 

\setcounter{proposition}{0}
\setcounter{equation}{0}
\propappc

\begin{appendix}

\section{Associative product of supertrace-free, graded matrices}

Let $\{ E_{A} \vert E_{A}\in\GSlg{n}{m}{0}, A=1,\cdots,n^{2}+m^{2}-1;
E_{A}\in\GSlg{n}{m}{1}, A=n^{2}+m^{2},\cdots,(n+m)^{2}-1 \}$ be a homogeneous basis of
$\GSl{n}{m}, n\neq m$. Our aim of this appendix is to investigate the associative
product of the homogeneous matrices $E_{A}$ in a similar way as it was done in
\cite{Macfarlane1} for trace-free, hermitian matrices.

If we introduce a graded anticommutator
\eqnappc
\BG
\label{a1}
\AComug{M}{M'} := MM' + (-1)^{\Over{M}\Over{M}'}M'M
\EG
of two homogeneous $M,M'\in\GMat{n}{m}$ we find the relations
\eqnappc
\BGG
\label{a2}
\Comug{M}{\Comug{M'}{M''}} -
\AComug{M}{\AComug{M'}{M''}} +
(-1)^{\Over{M}'\Over{M}''}\AComug{\AComug{M}{M''}}{M'}
&=& 0
\nonumber
\\
\Comug{\AComug{M}{M'}}{M''} -
\AComug{M}{\Comug{M'}{M''}} -
(-1)^{\Over{M}'\Over{M}''}\AComug{\Comug{M}{M''}}{M'}
&=& 0 
\EGG
between the graded commutator and the graded anticommutator of homogeneous 
$M,M',M''\in\GMat{n}{m}$.\newline
Because $\{ E_{A},1_{n+m}\}_{A=1,\cdots,(n+m)^{2}-1}$ forms a homogeneous basis of
$\GMat{n}{m}$ the graded anticommutator of $E_{A}$ and $E_{B}$ can be written according to
\eqnappc
\BG
\label{a3}
\AComug{E_{A}}{E_{B}} = 
\sum_{C=1}^{(n+m)^{2}-1}d_{AB}^{C}E_{C} + g_{AB}1_{n+m}
\EG
with uniquely determined coefficients $d_{AB}^{C}, g_{AB}\in\CC$. Then the associative
product of $E_{A}$ and $E_{B}$ is given by
\eqnappc
\BG
\label{a4}
E_{A}E_{B} = \frac{1}{2}\sum_{C=1}^{(n+m)^{2}-1}(c_{AB}^{C}+d_{AB}^{C})E_{C} +
\frac{1}{2}g_{AB}1_{n+m} \, .
\EG
Independent of the specific choice of the homogeneous basis
$\{ E_{A}\}_{A=1,\cdots,(n+m)^{2}-1}$ there exist a lot of relations between the
``structure constants'' $c_{AB}^{C}, d_{AB}^{C}$ and $g_{AB}$ which we summarize in
\begin{proposition}
\label{prop1a} 
$\mbox{}$ \newline
\vspace{-8mm}
\begin{enumerate}
\item[$\Rom{(i.)}$]
$c_{AB}^{C}$ and $d_{AB}^{C}$ vanish if $\Over{E}_{A}+\Over{E}_{B}+\Over{E}_{C}=\Over{1}$
and $g_{AB}$ vanishes if $\Over{E}_{A}+\Over{E}_{B}=\Over{1}$.
\item[$\Rom{(ii.)}$]
{\abovedisplayskip-3mm
\eqnappc
\BGG
\label{a5}
\sum_{B=1}^{(n+m)^{2}-1}(-1)^{\Over{E}_{B}}c_{AB}^{B} &=& 0
\nonumber
\\
\sum_{B=1}^{(n+m)^{2}-1}(-1)^{\Over{E}_{B}}d_{AB}^{B} &=& 0
\EGG}
\item[$\Rom{(iii.)}$]
$c_{ABC}$ and $d_{ABC}$, defined via
{\abovedisplayskip-1mm
\eqnappc
\BGG
\label{a6}
c_{ABC} &:=& \sum_{D=1}^{(n+m)^{2}-1}c_{AB}^{D}g_{DC}
\nonumber
\\
d_{ABC} &:=& \sum_{D=1}^{(n+m)^{2}-1}d_{AB}^{D}g_{DC} \, ,
\EGG}
are totally antisymmetric respectively totally symmetric in the $\ZZg$-graded sense.
\item[$\Rom{(iv.)}$]
{\abovedisplayskip-3mm
\eqnappc
\BGG
\label{a7}
\sum_{E=1}^{(n+m)^{2}-1}\lbr(-1)^{\Over{E}_{A}\Over{E}_{C}}c_{BC}^{E}c_{AE}^{D}+
(-1)^{\Over{E}_{B}\Over{E}_{A}}c_{CA}^{E}c_{BE}^{D}+
(-1)^{\Over{E}_{C}\Over{E}_{B}}c_{AB}^{E}c_{CE}^{D}\rbr = 0
\nonumber
\\
\sum_{E=1}^{(n+m)^{2}-1}\lbr c_{BC}^{E}c_{AE}^{D}-d^{E}_{AB}d_{EC}^{D}+
(-1)^{\Over{E}_{A}\Over{E}_{C}+\Over{E}_{B}\Over{E}_{C}}d_{CA}^{E}d_{EB}^{D}\rbr +
\qquad \qquad
\\
+ 2(-1)^{\Over{E}_{A}\Over{E}_{C}+\Over{E}_{B}\Over{E}_{C}}g_{CA}\delta_{B}^{D} -
2g_{AB}\delta_{C}^{D} = 0
\quad
\nonumber
\\
\sum_{E=1}^{(n+m)^{2}-1}\lbr d_{AB}^{E}c_{EC}^{D}-c_{BC}^{E}d_{AE}^{D}-
(-1)^{\Over{E}_{B}\Over{E}_{C}}c_{AC}^{E}d_{EB}^{D}\rbr = 0
\qquad \qquad
\nonumber
\EGG}
\item[$\Rom{(v.)}$]
If $K_{AB}:=K(E_{A},E_{B})$, where $K$ is the Killing form of $\GSl{n}{m}$, then
\eqnappc
\BGG
\label{a8}
K_{AB} = (n-m)^{2}g_{AB} =
\sum_{C,D=1}^{(n+m)^{2}-1}(-1)^{\Over{E}_{C}}c_{AD}^{C}c_{BC}^{D} =
\qquad \qquad \quad
\nonumber
\\
= \frac{(n-m)^{2}}{(n-m)^{2}-4}
\sum_{C,D=1}^{(n+m)^{2}-1}(-1)^{\Over{E}_{C}}d_{AD}^{C}d_{BC}^{D} \, .
\EGG
\item[$\Rom{(vi.)}$]
Denoting by $g^{AB}$ the components of the matrix inverse to $(g_{AB})$, then
\eqnappc
\BGG
\label{a9}
\sum_{B,C=1}^{(n+m)^{2}-1}g^{BC}c_{BC}^{A} &=& 0
\nonumber
\\
\sum_{B,C=1}^{(n+m)^{2}-1}g^{BC}d_{BC}^{A} &=& 0 \, .
\EGG
\item[$\Rom{(vii.)}$]
{\abovedisplayskip-3mm
\eqnappc
\BGG
\label{a10}
\sum_{C,D,E=1}^{(n+m)^{2}-1}g^{CD}c_{CE}^{A}c_{DB}^{E} &=&
(n-m)^{2}\delta^{A}_{B}
\nonumber
\\
\sum_{C,D,E=1}^{(n+m)^{2}-1}g^{CD}c_{CE}^{A}d_{DB}^{E} &=& 0
\\
\sum_{C,D,E=1}^{(n+m)^{2}-1}g^{CD}d_{CE}^{A}d_{DB}^{E} &=&
\lp(n-m)^{2}-4\rp\delta^{A}_{B}
\nonumber
\EGG}
\item[$\Rom{(viii.)}$]
{\abovedisplayskip-3mm
\eqnappc
\BGG
\label{a11}
\sum_{D,E,F,G=1}^{(n+m)^{2}-1}
(-1)^{\Over{E}_{A}\Over{E}_{E}}g^{DE}c_{EB}^{F}c_{AF}^{G}c_{DG}^{C} &=& 
\frac{1}{2}(n-m)^{2}c_{AB}^{C}
\nonumber
\\
\sum_{D,E,F,G=1}^{(n+m)^{2}-1}
(-1)^{\Over{E}_{A}\Over{E}_{E}}g^{DE}c_{EB}^{F}c_{AF}^{G}d_{DG}^{C} &=& 
-\frac{1}{2}(n-m)^{2}d_{AB}^{C}
\nonumber
\\
\sum_{D,E,F,G=1}^{(n+m)^{2}-1}
(-1)^{\Over{E}_{A}\Over{E}_{E}}g^{DE}c_{EB}^{F}d_{AF}^{G}d_{DG}^{C} &=& 
-\frac{1}{2}\lp(n-m)^{2}-4\rp c_{AB}^{C}
\\
\sum_{D,E,F,G=1}^{(n+m)^{2}-1}
(-1)^{\Over{E}_{A}\Over{E}_{E}}g^{DE}d_{EB}^{F}d_{AF}^{G}d_{DG}^{C} &=& 
\frac{1}{2}\lp(n-m)^{2}-12\rp d_{AB}^{C}
\nonumber
\EGG}
\end{enumerate}
\end{proposition}
${\sl Proof}$: (i.) is a reformulation of the homogeneity of $\{ E_{A}\}$. The first line of
(\ref{a8}) as well as (iii.) result from $K_{AB}=2(n-m)\Rom{Tr}_{s}(E_{A}E_{B})$. (ii.) is
a consequence of $\Rom{Tr}_{s}(\Rom{ad}E_{A})=0$ and of
$\sum_{B,C}\Rom{Tr}_{s}(g^{BC}\AComug{E_{B}}{E_{C}}E_{A})=0$.
(iv.) is a reformulation of the graded Jacobi identity and (\ref{a2}). Using the
second equation (\ref{a7}) one deduces the second line of (\ref{a8}). (vi.) follows from
(ii.) and (iii.). The left hand side of the first equation (\ref{a10}) is essentially
the second-order Casimir operator of $\GSl{n}{m}$ in the adjoint representation. The second
part of (\ref{a10}) follows from (iii.), whereas the third equation is a consequence of
the first part together with (iv.) and (vi.). The relations (viii.) are results of
calculations using (iii.),(iv.),(vi.) and (vii.).
\hfill $\Box$ 

\end{appendix}

\section*{Acknowledgment}

The authors would like to thank W.$\,$Bulla, J.$\,$Madore, P.$\,$Pre\v{s}najder and 
L.$\,$Pittner for helpful discussions and the ``Fonds zur F\"orderung der wissenschaftlichen Forschung (FWF)'' for support funding.


\begin{thebibliography}{99}

\bibitem{Bartocci1}
C.Bartocci, U.Bruzzo, D.Hern\'{a}ndez Ruip\'{e}rez.
{\em The Geometry of Supermanifolds}.\newline
Klu\-wer, 1991.

\bibitem{Connes1}
A.Connes.
{\em Noncommutative Geometry}.
Academic Press, 1994.

\bibitem{Coquereaux1}
R.Coquereaux, D.Kastler.
{\em Remarks on the Differential Envelopes of Associative Algebras}.
Pacif.Jour.Math. 137:245, 1989.

\bibitem{Cornwell1}
J.F.Cornwell.
{\em Group Theory in Physics III, Su\-per\-sym\-me\-tries and 
In\-fi\-nite-Di\-men\-sio\-nal 
Algebras}.
Academic Press, 1989.

\bibitem{Djemai1}
A.E.F.Djemai.
{\em Introduction to Dubois-Violette's Noncommutative Differential Geometry}.
Int.Jour.Theor.Phys. 34:801, 1995.

\bibitem{Dubois-Violette1}
M.Dubois-Violette, R.Kerner, J.Madore.
{\em Noncommutative differential geometry of matrix algebras}.
J.Math.Phys. 31(2):316, 1990.

\bibitem{Dubois-Violette10}
M.Dubois-Violette, R.Kerner, J.Madore.
{\em Noncommutative differential geometry and new models of gauge 
theory}.
J.Math.Phys. 31(2):323, 1990.

\bibitem{Dubois-Violette2}
M.Dubois-Violette, R.Kerner, J.Madore.
{\em Super Matrix Geometry}.
Class.Quant.Grav. 8:\newline
1077, 1991.

\bibitem{Dubois-Violette5}
M.Dubois-Violette.
{\em Derivations et calcul differentiel non commutatif}.
C.R.Acad.Sci.\newline
Paris I~307:403, 1988.

\bibitem{Frohlich1}
J.Fr\"{o}hlich, O.Grandjean, A.Recknagel.
{\em Supersymmetric Quantum Theory and (Non-Com\-mu\-ta\-ti\-ve) Differential
Geometry}.
Preprint ETH-TH/96-45.

\bibitem{Fuks1}
D.B.Fuks.
{\em Cohomology of Infinite-Dimensional Lie Algebras}.
Consultants Bureau, 1986.

\bibitem{Fuks2}
D.B.Fuks, D.A.Leites.
{\em Cohomology of Lie Superalgebras}.
Com.rend.Acad.bulg. 37:1595, 1984.

\bibitem{Jadczyk1}
A.Jadczyk, D.Kastler.
{\em Graded Lie-Cartan pairs I}.
Rep.Math.Phys. 25:1, 1987.

\bibitem{Gracia-Bondia1}
J.M.Gracia-Bondia, J.C.V\'{a}rilly.
{\em Connes' noncommutative differential geometry and the Standard Model}.
Jour.Geom.Phys. 12:223; 1993.

\bibitem{Grosse1}
H.Grosse, C.Klim\v{c}ik, P.Pre\v{s}najder.
{\em Field Theory on a Supersymmetric Lattice}.
Comm.\newline
Math.Phys. 185:155, 1997.

\bibitem{Grosse15}
H.Grosse, G.Reiter.
{\em The Fuzzy Supersphere}.
Jour.Geom.Phys. 28:349, 1998.

\bibitem{Hernandez1}
D.Hern\'{a}ndez Ruip\'{e}rez, J.Mu\~{n}oz Masqu\'{e}.
{\em Global Variational Calculus on Graded Manifolds, I: Graded Jet Bundles, Structure
1-Form and Graded Infinitesimal Contact Transformations}.
Jour.Math.$\,$pures et appl. 63:283, 1984.

\bibitem{Hernandez2}
D.Hern\'{a}ndez Ruip\'{e}rez, J.Mu\~{n}oz Masqu\'{e}.
{\em Construction intrins\`{e}que du faisceau de Berezin d'une vari\'{e}t\'{e} gradu\'{e}e}.
C.R.Acad.Sci.Paris 301:915, 1985.

\bibitem{Kalau1}
W.Kalau, M.Walze.
{\em Supersymmetry and noncommutative geometry}.
Jour.Geom.Phys. 22:\newline
77, 1997.

\bibitem{Kastler1}
D.Kastler.
{\em Cyclic cohomology within the differential envelope}.
Hermann, 1988.

\bibitem{Kerner1}
R.Kerner.
{\em Graded non-commutative geometries}.
Jour.Geom.Phys. 11:325, 1993.

\bibitem{Kostant1}
B.Kostant.
{\em Graded manifolds, graded Lie theory, and prequantization}.
In: Lecture Notes in Mathematics 570:177, 1977.

\bibitem{Macfarlane1}
A.J.Macfarlane, A.Sudbery, P.H.Weisz.
{\em On Gell-Mann's $\lambda$-Matrices, d- and f-Tensors, Octets, and
Parametrizations of SU(3)}.
Comm.Math.Phys. 11:77, 1968.

\bibitem{Madore10}
J.Madore.
{\em An Introduction to Noncommutative Differential Geometry 
and its Physical Applications}.
Cambridge University Press, 1995.

\bibitem{Nastasescu1}
C.N\v{a}st\v{a}secu, F.van Oystaeyen.
{\em Graded Ring Theory}.
North-Holland, 1982.

\bibitem{Pierce1}
R.S.Pierce.
{\em Associative Algebras}.
Springer, 1982.

\bibitem{Pittner1}
L.Pittner.
{\em Algebraic Foundations of Non-Commutative Differential Geometry
and Quantum Groups}.
Springer, 1996.

\bibitem{Scheunert8}
M.Scheunert, R.B.Zhang.
{\em Cohomology of Lie superalgebras and of their generalizations}.
J.Math.Phys. 39:5024, 1998.

\end{thebibliography}
\end{document}